\documentclass[preprint]{elsarticle}
\usepackage{amsmath}
\usepackage{graphicx}
\usepackage{hyperref}
\usepackage{color}

\begin{document}
\begin{frontmatter}

\title{Comparison between Fermion Bag Approach and Complex Langevin Dynamics
for Massive Thirring Model at Finite Density in 0 + 1 Dimensions}

\author{Daming Li\corref{cor1}}
\ead{lidaming@sjtu.edu.cn}
\address{School of Mathematical Sciences, Shanghai Jiao Tong University, Shanghai, 200240, China}

\cortext[cor1]{Corresponding author}

\begin{abstract}
We consider the massive Thirring model at finite density in 0+1
dimension. The fermion bag approach, Langevin dynamics and complex
Langevin dynamics are adopted to attack the sign problem for this
model. Compared with the complex Langevin dynamics, both fermion bag
approach and Langvin dynamics avoid the sign problem. The fermion
density and chiral condensate, which are obtained by these numerical
methods, are compared with the exact results. The advantages of the
fermion bag approach over the other numerical methods are also
discussed.
\end{abstract}


%\pacs{ 05.50.+q, %Lattice theory and statistics
%% 05.30.Rt, %Quantum phase transitions
%% 03.70.+k % Theory of quantized fields
% 71.10.Fd, % Lattice fermion models
% 02.70.Ss, %Quantum Monte Carlo methods%
%% 11.30.Rd % Chiral symmetries
%}

\end{frontmatter}

\section{Introduction}
The usual sampling methods, e.g., Langevin dynamics and Monte Carlo
method, fail for the complex action, since the Boltzmann factor can
not be regarded as the probability density. This problem, called the
sign problem,  is still the main obstacle of the computation in
lattice quantum field. Three reasons will always lead to the complex
action: (1) grand partition function with finite density; (2)
fermion systems; (3) topological terms in the action.

To overcome the sign problem, the complex Langevin dynamics, which
is obtained from the complexification of the Langevin dynamics, was
used, which is rather successful in XY model \cite{Aarts_2010_0617},
Bose gas \cite{Aarts_2009_2089}, Thirring model
\cite{Pawlowski_2013_2249}, Abelian and
 Non-Abelian lattice gauge model \cite{Flower_1986},
QCD model \cite{Sexty_2014_7748}, and its simplified model including
one link U(1) model, one link SU(3) model, QCD model in the heavy
mass limit \cite{Aarts_2008_1597}, one link SU(N) model
\cite{Aarts_1212.5231}, SU(3) spin model \cite{Aarts_2012_4655},
Polykov chain model \cite{Aarts_2013_6425}. For some range of
chemical potential $\mu$ and large fluctuation, the complex Langevin
may fails, e.g., the XY model at finite chemical potential for small
$\beta$ (large fluctuation) \cite{Aarts_2010_3468} and in the
Thirring model in 0+1 dimension \cite{Pawlowski_2013_094503}. Unfortunately from early
studies of complex Langevin evolutions
\cite{Hamber_330}\cite{Flower_330}\cite{Ilgenfritz_327} until this
day, the convergence properties of complex Langevin equations are
not well understood. Recently Aarts etc. provided a criterion for
checking the correctness of the complex Langevin dynamics
\cite{Aarts_2011_3270}.

It is possible to find suitable variables to  represent the
partition function such that the action is real. This is called the
dual variable method. It is successfully applied to many models,
including Bose gas \cite{Gattringer_2013}, SU(3) spin model
\cite{Mercado_2012}, U(1) and Z(3) gauge Higgs lattice theory
\cite{Mercado_2013}, massive lattice Schwinger model
\cite{Gattringer_2015}, O(3), O(N) and CP(N-1) model
\cite{Bruckmann_05482}\cite{Bruckmann_2015}\cite{Wolff_0908.0284}\cite{Wolff_1001.2231},
fermion bag approach \cite{Chandrasekharan_1304.4900},
 4-fermion lattice theory, including massless Thirring model
  \cite{Chandrasekharan_2011_5276}, Gross-Neveu model
  \cite{Chandrasekharan_2012_6572}, Yukawa model
  \cite{Chandrasekharan_1205.0084},
  Non-Abelian Yang-Mills model \cite{Oeckl_0008095}\cite{Cherrington_0705.2629}, and its coupling with fermion field
  \cite{Cherrington_0710.0323},
lattice chiral model, and Sigma model \cite{Pfeiffer_033501}.
Although the routine to find the dual variable is case by case, it's
technique behind is based on the high temperature expansion. For the
fermion system, the dual method is called fermion bag approach
\cite{Chandrasekharan_1304.4900}. This numerical method not only
overcome the sign problem for model with small chemical potential,
but also a high computational efficiency is achieved for the small
or large coupling between the fermions. In this paper we compare the
Langvin dynamics, the complex Langevin dynamics and the fermion bag
approach for the massive Thirring model at finite density in 0+1
dimension. The chiral condensate and fermion density obtained by
these two numerical methods, are compared with the exact result.

The arrangement of the paper is as follows. In section \ref{model},
the Fermion bag approach for Thirring model is presented. In section
\ref{Langevin dynamics}, the complex Langevin dynamics and (real)
Langevin dynamics are given for this model by introducing a bosonic
variable. In section \ref{results}, the implementation of these
numerical methods are presented, and are compared with the exact
result. Conclusions are given in section \ref{conclusion}.

\section{Thirring model in 0+1 dimension}\label{model}
The lattice partition function for the massive Thirring model at the
finite density in 0+1 dimension reads
\begin{eqnarray}\label{2016_4_27_0}
 Z = \int d\bar\psi d\psi e^{-S}
\end{eqnarray}
where $d\bar\psi d\psi = \prod_{t=0}^{N-1} d\bar\psi_t d\psi_t$ is
the measure of the Grassmann fields $\psi = \{\psi_t\}_{t=0}^{N-1}$
and $\bar\psi=\{\bar\psi_t\}_{t=0}^{N-1}$, with even number of sites
$N$. We adopt anti-periodic condition for $\psi$ and $\bar\psi$
$$ \psi_N = -\psi_0, \quad \psi_{-1} = -\psi_{N-1}, \quad
\bar\psi_{N} = -\bar\psi_0, \quad \bar\psi_{-1} = -\bar\psi_{N-1}
$$
The staggered fermion action $S$ is
\begin{eqnarray}\label{2016_5_13_1}
S = \sum_{t,\tau=0}^{N-1} \bar\psi_tD_{t,\tau}\psi_\tau - U
\sum_{t=0}^{N-1} \bar\psi_t\psi_t\bar\psi_{t+1}\psi_{t+1}
\end{eqnarray}
  Here $t+1 $ is always understood to be $(t+1)\mod N$, which becomes 0 if
  $t=N-1$
since the period is $N$. The coupling constant $U$ between fermions
is nonnegative. The fermion matrix $D$ is
$$ D_{t,\tau} = D(\mu,m)_{t,\tau} = \frac{1}{2}\Big(
s^1_t  e^{\mu  }\delta_{t+1,\tau}- s^2_t e^{-\mu
}\delta_{t,\tau+1}\Big)+ m\delta_{t,\tau}, \quad 0\leq t,\tau\leq
N-1
$$
with the chemical potential $\mu$ and fermion mass $m$. The
antiperiodic condition for the $\psi$ and $\bar \psi$ are accounted
for by the sign $s^1$ and $s^2$
\begin{eqnarray}
 s^1_t =  \left\{
  \begin{array}{l l}
-1  & \quad \text{if \ $t=N-1$}\\
1 &  \quad \text{if \ $0\leq t< N-1$}\\
   \end{array} \right. , \quad
 s^2_t =  \left\{
  \begin{array}{l l}
-1  & \quad \text{if \ $t=0$}\\
1 &  \quad \text{if \ $1\leq t\leq N-1$}\\
   \end{array} \right.
\end{eqnarray}
A periodic extension for $s^1$ and $s^2$ is used to define them for
any $t$. Thus we have $s^1_t = s^2_{t+1}$ for any $t$. By using this
formula for $D$, the action $S$ in (\ref{2016_5_13_1}) can be
written as a sum over edges $(t,t+1)$
\begin{eqnarray*}
S &=& \frac{1}{2} e^\mu (\bar\psi_0\psi_1+
\cdots+\bar\psi_{N-2}\psi_{N-1}+\bar\psi_{N-1}\psi_0)
\\ && + \frac{1}{2} e^{-\mu} (\psi_{N-1}\bar\psi_0+\psi_0\bar\psi_1+
\cdots+\psi_{N-2}\bar\psi_{N-1})  \\
&&  +  m \sum_{t=0}^{N-1} \bar\psi_t\psi_t + U \sum_{t=0}^{N-1}
(\bar\psi_t\psi_{t+1}) (\bar\psi_{t+1}\psi_t)
\end{eqnarray*}

It is easy to check the symmetry of the fermion matrix $D(\mu,m)$
with respect to $\mu$ and $m$
\begin{eqnarray}\label{2016_4_27_4}
D(\mu,m)_{t,\tau}=-D(-\mu,-m)_{\tau, t} \Longrightarrow \det
D(\mu,m) = \det D(-\mu,-m)
 \end{eqnarray}
\begin{eqnarray}\label{2016_4_27_5}
 \varepsilon   D(\mu,m) \varepsilon  = -  D(\mu,-m) \Longrightarrow \det
D(\mu,m) = \det D(\mu,-m)
\end{eqnarray}
where $\varepsilon_{t,\tau}= \delta_{t,\tau}\varepsilon_{t}$, $
\varepsilon_{t}=(-1)^t$. Thus it is enough to study $\mu\geq 0$ and
$m\geq 0$ in the massive Thirring model.

The idea of the fermion bag approach is to expand the Boltzmann
factor $e^{-S}$ by the high temperature expansion,
\begin{eqnarray}
\exp(-S) & = & \exp\Big(-\sum_{t,\tau=0}^{N-1}
\bar\psi_tD_{t,\tau}\psi_\tau\Big)\prod_{t=0}^{N-1}
\exp\Big(U \bar\psi_t\psi_t\bar\psi_{t+1}\psi_{t+1}\Big)\nonumber\\
& = & \exp\Big(-\sum_{t,\tau=0}^{N-1}
\bar\psi_tD_{t,\tau}\psi_\tau\Big)\prod_{t=0}^{N-1}\sum_{k_t=0}^1(U\bar\psi_t\psi_t\bar\psi_{t+1}\psi_{t+1}
)^{k_t}
\end{eqnarray}
and the partition function $Z$ is written as the sum over the
configuration $k=(k_t=0,1)_{t=0}^{N-1}$
\begin{eqnarray}\label{2016_4_27_2} Z = \sum_{k}   U^{j} C(t_1,\cdots,t_{2j})
\end{eqnarray}
  where $k_{t}=0,1$ for all two neighboring sites $(t,t+1)$, which
  satisfies
$ k_{t-1} + k_t\leq 1$ for all site $t$. If $k_t=1$, we say there is
a bond connecting $t$ and $t+1$; otherwise, there are no bonds
connecting them. For given configuration $k$, for example, there are
$j$ bonds $(t_1,t_2),\cdots,(t_{2j-1},t_{2j})$ connecting $2j$
different sites $(t_1,\cdots,t_{2j})$. For such kind of
configuration $k$, $C$ in (\ref{2016_4_27_2}) is given by
\begin{eqnarray}\label{2016_4_27_6}  C(t_1,\cdots,t_{2j}) &=& \int d\bar\psi d\psi \exp\Big(-\sum_{t,\tau=0}^{N-1}
\bar\psi_tD_{t,\tau}\psi_\tau\Big) \bar\psi_{t_1}\psi_{t_1}\cdots
\bar\psi_{t_{2j}}\psi_{t_{2j}}   \\
 & = &  \det D\det
G(\{t_1,\cdots,t_{2j}\})    =  \det D(\backslash
\{t_1,\cdots,t_{2j}\}) \nonumber
\end{eqnarray}
where $G(\{t_1,\cdots,t_{2j}\})$ is a $(2j)\times (2j)$ matrix of
propagators between $2j$ sites $t_i$, $i=1,\cdots,2j$, whose matrix
element are $G(\{t_1,\cdots,t_{2j}\})_{i,l}=D^{-1}_{{t_{i}},t_{l}}$,
$i,l=1,\cdots,2j$. The matrix $G(\{t_1,\cdots,t_{2j}\})$ depends on
the order of $\{t_1,\cdots,t_{2j}\}$, but it's determinant does not.
$D(\backslash \{t_1,\cdots,t_{2j}\})$ is the $(N-2j)\times (N-2j)$
matrix which is obtained by deleting rows and columns corresponding
to sites $t_1, \cdots, t_{2j}$.  Thus if $j$ is small, we use
$G(\{t_1,\cdots,t_{2j}\})$ to calculate $C(t_1,\cdots,t_{2j})$;
otherwise, we adopted $D(\backslash \{t_1,\cdots,t_{2j}\})$ to
calculate $C(t_1,\cdots,t_{2j})$. Because of the symmetry
(\ref{2016_4_27_4}) and (\ref{2016_4_27_5}) of $D$, it is easy to show  
that for any $\mu$, $m$ and any number of different sites $t_1,
\cdots, t_n$
\begin{eqnarray}\label{2016_10_1_1}
&&   C(t_1,\cdots,t_{n}; D(\mu,m)) =
 (-1)^n C(t_1,\cdots,t_{n}; D(\mu,-m)) \nonumber \\ &  =  &  C(t_1,\cdots,t_{n};
  D(-\mu,m)) = (-1)^n C(t_1,\cdots,t_{n};
  D(-\mu,-m))
\end{eqnarray}
 where $C(t_1,\cdots,t_{n}; D(\mu,m))$ denote the function $C$ in
(\ref{2016_4_27_6}) since it depends on the fermion matrix
$D(\mu,m)$. We can rigorously prove that for any $\mu\geq 0$ and $m>
0$, $C(t_1,\cdots,t_j)>0$ for any number of different sites
$\{t_i\}_{i=1}^j$ (\ref{Appendix_1}). Thus the sign problem is avoided by the
presentation (\ref{2016_4_27_2}) of the partition function for the
massive Thirring model with finite density in 0+1 dimensions.

The chiral condensate is
\begin{eqnarray}\label{2016_4_27_12}
\langle \bar \psi \psi \rangle  = \frac{1}{N}\frac{\partial \ln
Z}{\partial m} = \frac{1}{N}\Big\langle \frac{ \partial_m
C(t_1,\cdots,t_{2j})}{ C(t_1,\cdots,t_{2j})}\Big  \rangle
\end{eqnarray}
where the average is taken with respect to the weight of the
partition function (\ref{2016_4_27_2}). From the first line in (\ref{2016_10_1_1}), $C(t_1,\cdots,t_{2j})$ is even in $m$ and the chiral condensate
$\langle \bar \psi \psi \rangle$ vanishes if $m=0$. Similar to the calculation
of $C$ in (\ref{2016_4_27_6}), we also have two methods to calculate
its partial derivative $\partial_m C$
\begin{eqnarray}\label{2016_5_3_4}
\partial_m C(t_1,\cdots,t_{2j}) &=&  \sum_{t\neq t_1,\cdots,t_{2j}} \det D\det
G(\{t, t_1,\cdots,t_{2j}\})  \nonumber \\ & = &  \sum_{t\neq
t_1,\cdots,t_{2j}} \det D(\backslash \{t, t_1,\cdots,t_{2j}\})
\end{eqnarray}

The fermion density is
\begin{eqnarray}\label{2016_4_27_14}
\langle n \rangle  = \frac{1}{N}\frac{\partial \ln Z}{\partial \mu}
= \frac{1}{N}\Big\langle \frac{ \partial_\mu C(t_1,\cdots,t_{2j})}{
C(t_1,\cdots,t_{2j})}\Big  \rangle
\end{eqnarray}
which vanishes if $\mu=0$ since $C(t_1,\cdots,t_{2j})$ is even in $\mu$ by (\ref{2016_10_1_1}).
The partial derivative of $C$ with respect to $\mu$ is
\begin{eqnarray*}
&& \partial_\mu  C(t_1,\cdots,t_{2j})\nonumber
\\& = &    \sum_{t,t+1\neq
t_1,\cdots,t_{2j}} \frac{1}{2} e^\mu s^1_{t} \det D  \det
G[(t_1,\cdots,t_{2j},t+1),(t_1,\cdots,t_{2j},t)] +\\ &&
\sum_{t,t-1\neq t_1,\cdots,t_{2j}}\frac{1}{2} e^{-\mu}s^2_{t}\det D
\det G[(t_1,\cdots,t_{2j},t-1),(t_1,\cdots,t_{2j},t)]
\end{eqnarray*}
where the $(2j+1)\times (2j+1)$ propagator matrix
$G[(t_1,\cdots,t_{2j},t\pm 1),(t_1,\cdots,t_{2j},t)]$ has $(i,l)$
matrix element $D^{-1}_{t_i,\tau_l}$, $i,l=1,\cdots,2j+1$. Here we
use notations $t_{2j+1}\equiv t\pm 1$, $\tau_l=t_l$,
$l=1,\cdots,2j$, $\tau_{2j+1}=t$. The ratio in (\ref{2016_4_27_14})
is
\begin{eqnarray*} && \frac{ \partial_\mu
C(t_1,\cdots,t_{2j})}{ C(t_1,\cdots,t_{2j})} \\ & = &
\frac{1}{2}e^\mu \sum_{t,t+1\neq t_1,\cdots,t_{2j}} s^1_{t}
\frac{\det G[(t_1,\cdots,t_{2j},t+1),(t_1,\cdots,t_{2j},t)]}{\det
G[(t_1,\cdots,t_{2j}),(t_1,\cdots,t_{2j})]} +  \\ && \frac{1}{2}
e^{-\mu}\sum_{t,t-1\neq t_1,\cdots,t_{2j}} s^2_{t} \frac{ \det
G[(t_1,\cdots,t_{2j},t-1),(t_1,\cdots,t_{2j},t)]}{\det
G[(t_1,\cdots,t_{2j}),(t_1,\cdots,t_{2j})]}
\end{eqnarray*}
Note that
$G[(t_1,\cdots,t_{2j}),(t_1,\cdots,t_{2j})]=G(\{t_1,\cdots,t_{2j}\})$.
The ratio between the determinant can be obtained by
\begin{eqnarray*}
&& \frac{\det G[(t_1,\cdots,t_{2j},t\pm
1),(t_1,\cdots,t_{2j},t)]}{\det
G[(t_1,\cdots,t_{2j}),(t_1,\cdots,t_{2j})]}  =
  G[(t\pm 1),(t)] \\ &&  - G[(t\pm 1),\textrm{occu\_sites}]G[\textrm{occu\_sites},\textrm{occu\_sites}]^{-1}
G[\textrm{occu\_sites},(t)]
\end{eqnarray*}
where $\textrm{occu\_sites}=(t_1,\cdots,t_{2j})$ denotes $2j$
occupied sites. $G[\textrm{occu\_sites},\textrm{occu\_sites}]$ is
the $2j\times 2j$ propagator matrix with $(i,l)$ matrix element
$D^{-1}_{t_i,t_l}$, $i,l=1,\cdots,2j$. Another form of $\partial_\mu
C$ is
\begin{eqnarray*}
&& \partial_\mu  C(t_1,\cdots,t_{2j})\\ & =  &  \frac{1}{2}e^\mu
\sum_{t,t+1\neq t_1,\cdots,t_{2j}}s^1_{t} (-1)^{(t)-(t+1)}
  \det
D[\backslash(t_1,\cdots,t_{2j},t),\backslash(t_1,\cdots,t_{2j},t+1) ]+ \\
&&   \frac{1}{2}e^{-\mu} \sum_{t,t-1\neq t_1,\cdots,t_{2j}} s^2_{t}
(-1)^{(t)-(t-1)}    \det
D[\backslash(t_1,\cdots,t_{2j},t),\backslash(t_1,\cdots,t_{2j},t-1)
]
\end{eqnarray*}
where $(t)$ and $(t+1)$ denote the position of $t$ and $t+1$ in the
$N-2j$ $\textrm{free\_sites}=\backslash \{t_1,\cdots,t_{2j}\}$,
respectively.
$D[\backslash(t_1,\cdots,t_{2j},t),\backslash(t_1,\cdots,t_{2j},t\pm
1) ]$ denotes the $(N-2j-1)\times (N-2j-1)$ matrix obtained from $D$
by deleting rows $(t_1,\cdots,t_{2j},t)$ and columns
$(t_1,\cdots,t_{2j},t\pm 1)$. Using this formula, one has
\begin{eqnarray*} && \frac{ \partial_\mu
C(t_1,\cdots,t_{2j})}{ C(t_1,\cdots,t_{2j})} \\ & = &
\frac{1}{2}e^\mu    \sum_{t,t+1\neq t_1,\cdots,t_{2j}} s^1_{t}
(-1)^{(t)-(t+1)} \frac{\det D[
\backslash(t_1,\cdots,t_{2j},t),\backslash(t_1,\cdots,t_{2j},t+1)
]}{ \det
D[\backslash(t_1,\cdots,t_{2j}),\backslash(t_1,\cdots,t_{2j}) ]} + \\
&&   \frac{1}{2}e^{-\mu} \sum_{t,t-1\neq t_1,\cdots,t_{2j}} s^2_{t}
(-1)^{(t)-(t-1)} \frac{ \det D[
\backslash(t_1,\cdots,t_{2j},t),\backslash(t_1,\cdots,t_{2j},t-1)
]}{ \det
D[\backslash(t_1,\cdots,t_{2j}),\backslash(t_1,\cdots,t_{2j}) ]}
\end{eqnarray*}
Note that
$D[\backslash(t_1,\cdots,t_{2j}),\backslash(t_1,\cdots,t_{2j}) ]=
D(\backslash \{t_1,\cdots,t_{2j}\})$. The ratio between the
determinant is
 \begin{eqnarray*}
  (-1)^{(t)-(t\pm 1)} \frac{ \det D[
\backslash(t_1,\cdots,t_{2j},t),\backslash(t_1,\cdots,t_{2j},t\pm 1)
]}{ \det
D[\backslash(t_1,\cdots,t_{2j}),\backslash(t_1,\cdots,t_{2j}) ]}  =
  \textrm{D\_Inv}[(t\pm 1),(t)]
\end{eqnarray*}
where $  \textrm{D\_Inv}[(t\pm 1),(t)]$ is the $(t\pm 1, t)$ matrix
element of $(N-2j)\times (N-2j)$ matrix $\textrm{D\_Inv}$, which is
the inverse of $(N-2j)\times (N-2j)$ matrix $D[
\textrm{free\_sites},\textrm{free\_sites}]$.

The Monte Carlo algorithm based on the partition function in
(\ref{2016_4_27_2}) can be found in
Ref.\cite{Chandrasekharan_2011_5276}. We adopt the following steps
to update the current configuration. Assume that the current
configuration $k$ has $n_b$ bonds
$$ C=([t_1,t_2],\cdots, [t_{2n_b-1},t_{2n_b}]) $$
Try to delete a bond, e.g. $[t_{2n_b-1},t_{2n_b}]$ from the current
configuration $C$ to be $$ C^\prime=([t_1,t_2],\cdots,
[t_{2n_b-3},t_{2n_b-2}])
$$
According to the detailed balance
\begin{eqnarray}\label{2016_4_27_8}
  W(C) P_{try}(C\rightarrow C^\prime) P_{acc}(C\rightarrow C^\prime) =
W(C^\prime) P_{try}(C^\prime\rightarrow C)
P_{acc}(C^\prime\rightarrow C)
\end{eqnarray}
where $W(C)$ and $W(C^\prime)$ are the weight in the partition
function (\ref{2016_4_27_2}) for the configuration $C$ and
$C^\prime$, respectively. The try probability from $C (C^\prime)$ to
$C^\prime (C)$ are
\begin{eqnarray*} P_{try}(C\rightarrow C^\prime) = \frac{1}{n_b}, \quad
P_{try}(C^\prime\rightarrow C) = \frac{1}{n_f}
\end{eqnarray*}
respectively.  Here $n_f$ is the number of bonds which can be
created from the configuration $C^\prime$. Thus the accept
probability from $C$ to $C^\prime$ is
\begin{eqnarray*}
 P_{acc}(C\rightarrow C^\prime) = \frac{n_b}{n_f} \frac{W(C^\prime)}{W(C)}
\end{eqnarray*}
Try to add a bond, e.g. $ [t_{2n_b+1},t_{2n_b+2}] $ from the current
configuration $C$ to be
\begin{eqnarray*}
 C^\prime=([t_1,t_2],\cdots, [t_{2n_b-1},t_{2n_b}],[t_{2n_b+1},t_{2n_b+2}])
\end{eqnarray*}
The detailed balance is Eq. (\ref{2016_4_27_8}) where
\begin{eqnarray*}
P_{try}(C\rightarrow C^\prime) = \frac{1}{n_f}, \quad
P_{try}(C^\prime\rightarrow C) = \frac{1}{n_b+1}
\end{eqnarray*}
 Here $n_f$ is the
number of bonds which can be created from the configuration $C$.
Thus the accept probability from $C$ to $C^\prime$ is
\begin{eqnarray*}
 P_{acc}(C\rightarrow C^\prime) = \frac{n_f}{n_b+1} \frac{W(C^\prime)}{W(C)}
\end{eqnarray*}
Try to delete a bond, e.g. $ [t_{2n_b-1},t_{2n_b}] $  from the
current configuration $C$ and then add a bond, e.g., $
[y_{2n_b-1},y_{2n_b}] $ to be
$$ C^\prime=([t_1,t_2],\cdots, [t_{2n_b-3},t_{2n_b-2}],[y_{2n_b-1},y_{2n_b}])
$$
In the detailed balance (\ref{2016_4_27_8}),
\begin{eqnarray*}  P_{try}(C\rightarrow C^\prime) = P_{try}(C^\prime\rightarrow C) =
\frac{1}{n_bn_f}
\end{eqnarray*} Here $n_f$ is the number of bonds which can be
created from the configuration $C$ where $ [t_{2n_b-1},t_{2n_b}] $
is deleted. Thus the accept probability to move a bond is
\begin{eqnarray*}
 P_{acc}(C\rightarrow C^\prime) =  \frac{W(C^\prime)}{W(C)}
\end{eqnarray*}

\section{Complex Langevin dynamics}\label{Langevin dynamics}
The Grassmann fields $\bar \psi$ and $\psi$ can be eliminated if an
bosonic variable $A_t$ are introduced
\begin{eqnarray}\label{2016_5_4_1}
   &&\exp\Big(U \sum_{t=0}^{N-1}
\bar\psi_t\psi_t\bar\psi_{t+1}\psi_{t+1}\Big)  =\int
\prod_{t=0}^{N-1}dA_t
\exp\Big(-\frac{1}{8U}\sum_{t=0}^{N-1}A_{t}^2\Big) \nonumber \\
&& \exp\Big(-\sum_{t,\tau=0}^{N-1} \bar\psi_t  \frac{i}{2}(e^{\mu
}A_t \delta_{t+1,\tau}+e^{-\mu}A_\tau \delta_{t,\tau+1}) \psi_\tau
\Big)
\end{eqnarray}
where we omit a term depending on $U$. Inserting this formula to the
partition function $Z$ in (\ref{2016_4_27_0}) and integrating over
the Grassmann fields,  one has
\begin{eqnarray}\label{2016_4_27_10}
 Z  =
\int \prod_{t=0}^{N-1}dA_t \exp\Big(-\frac{1}{8U}\sum_{t=0}^{N-1}
A_t^2   \Big)\det K = \int \prod_{t=0}^{N-1}dA_t
e^{-S_{\textrm{eff}}}
\end{eqnarray}
where the $N\times N$ fermion matrix $K$ under the bosonic variable
$A$ is
\begin{eqnarray}\label{2016_5_13_4} K_{t,\tau} = \frac{1}{2}\Big(
(s^1_t+iA_t) e^{\mu  }\delta_{t+1,\tau}-(s^2_t-iA_\tau)e^{-\mu
}\delta_{t,\tau+1}\Big)+ m\delta_{t,\tau}
\end{eqnarray}
for $0\leq t,\tau\leq N-1$. The fermion matrix $K$ becomes the
fermion matrix $D$ if $A=0$. This form of partition function in
(\ref{2016_4_27_10}) was studied in
Ref.\cite{Pawlowski_2013_094503}. The effective action in
(\ref{2016_4_27_10}) is
\begin{eqnarray}\label{2016_5_4_16}
  S_{\textrm{eff}} = \frac{1}{8U}\sum_{t=0}^{N-1}
A_{t}^2  -    \ln \det K
\end{eqnarray}
  is complex. The discrete complex Langevin
dynamics for (\ref{2016_4_27_10})
\begin{eqnarray}\label{2016_5_2_1}
 A_{t,\Theta+\Delta \Theta} = A_{t, \Theta}   -\Delta \Theta \frac{\partial S_{\textrm{eff}}}{\partial
A_{t,\Theta}} + \sqrt{2\Delta \Theta} \eta_{t,\Theta}, \quad 0\leq
t\leq N-1, \quad \Theta = 0, 1,\cdots
\end{eqnarray}
where $\Theta$ denotes the discrete complex Langevin time, $\Delta
\Theta$ is the time step. The real white noise $\eta_{t,\Theta}\sim
{\cal N} (0,1)$ satisfies
$$
\langle \eta_{t,\Theta} \eta_{\tau,\Theta^\prime} \rangle =
\delta_{t,\tau}\delta_{\Theta,\Theta^\prime}
$$
Since $S_{\textrm{eff}}$ is complex,  $A_{t, \Theta}$ is also
complex. Using (\ref{2016_5_13_4}) and (\ref{2016_5_4_16}), the
drift force in (\ref{2016_5_2_1}) is
\begin{eqnarray}\label{2016_5_4_20}
-\frac{\partial S_{\textrm{eff}}}{\partial A_{t}}    = -\frac{1}{4U}
A_t + \frac{i}{2} \Big(  e^{\mu } K^{-1}_{t+1,t} +  e^{-\mu }
K^{-1}_{t,t+1} \Big)
\end{eqnarray}
The chiral condensate in (\ref{2016_4_27_12}) is written as
\begin{eqnarray} \label{2016_4_27_16}
 \langle \bar\chi\chi \rangle
= \frac{1}{N}\langle \textrm{Tr}(K^{-1} ) \rangle
\end{eqnarray}
and the fermion density in (\ref{2016_4_27_14}) reads
\begin{eqnarray}\label{2016_4_27_18}
 \langle n \rangle =
\frac{1}{N}\Big\langle \textrm{Tr}(K^{-1} \frac{\partial K}{\partial
\mu} ) \Big\rangle
\end{eqnarray}
Here the average is taken with respect to weight
$e^{-S_{\textrm{eff}}}$.

The equation (\ref{2016_5_4_1}) is also valid if $A_t$ and $A_\tau$
are replaced by $s^1_tA_t$, $s^2_tA_\tau$, respectively. The
partition function $Z$ is
\begin{eqnarray}\label{2016_5_4_2}
 Z  =
\int \prod_{t=0}^{N-1}dA_t \exp\Big(-\frac{1}{8U}\sum_{t=0}^{N-1}
A_t^2   \Big)\det \tilde K
\end{eqnarray}
where the $N\times N$ fermion matrix $\tilde K$ under the bosonic
variable $A$ is
\begin{eqnarray}\label{2016_5_4_6}
  \tilde K_{t,\tau} = \frac{1}{2}\Big(
s^1_t(1+iA_t) e^{\mu  }\delta_{t+1,\tau}-s^2_t(1-iA_\tau)e^{-\mu
}\delta_{t,\tau+1}\Big)+ m\delta_{t,\tau}
\end{eqnarray}
for $0\leq t,\tau\leq N-1$.   Fortunately the determinant of $\tilde
K$ can be calculated by \cite{Molinari.2221}
\begin{eqnarray}\label{2016_5_4_3}
\det \tilde K& =&      \frac{e^{N\mu}}{2^N}\prod_{t=0}^{N-1}
(1+iA_t) + \frac{e^{-N\mu}}{2^N}\prod_{t=0}^{N-1} (1-iA_t) +
\textrm{Tr} (T)
\end{eqnarray}
where the $2\times 2$ transfer matrix $T$ is
\begin{eqnarray}\label{2016_5_4_4}
T(A_0,\cdots,A_{N-1}) =  \left(
\begin{array}{cc}
m &  \frac{1}{4}(1+A_{0}^2) \\
1 & 0
\end{array} \right) \cdots \left(
\begin{array}{cc}
m &  \frac{1}{4}(1+A_{N-1}^2) \\
1 & 0
\end{array} \right)
\end{eqnarray}
Inserting (\ref{2016_5_4_3}) into the partition function $Z$ in
(\ref{2016_5_4_2}), one has $ Z  =  Z_1(\mu) + Z_2(m)$ where
\begin{eqnarray*}
 Z_1(\mu) = 2 (2\pi U)^{N/2}\cosh(N\mu), \quad
 Z_2(m) =
\int \prod_{t=0}^{N-1}dA_t \exp\Big(-\frac{1}{8U}\sum_{t=0}^{N-1}
A_t^2   \Big) \textrm{Tr} (T)
\end{eqnarray*}  The chiral condensate
is
\begin{eqnarray} \label{2016_5_4_10}
 \langle \bar\chi\chi \rangle
= \frac{1}{N} \frac{ \langle \textrm{Tr} (\partial_m T)
\rangle_0}{2^{-N}2\cosh(N\mu)+  \langle \textrm{Tr} (T)\rangle_0 }
\end{eqnarray}
where the average is taken with respect to the weight $\exp
(-\frac{1}{8U}\sum_{t=0}^{N-1} A_t^2 ) $. The fermion density reads
\begin{eqnarray}\label{2016_5_4_11}
 \langle n \rangle =
\frac{2 \sinh(N\mu)}{ 2 \cosh(N\mu)  +  2^{N}  \langle \textrm{Tr}
(T)\rangle_0 }
\end{eqnarray}   Since the observable
$\textrm{Tr} (T)>0$ and $\textrm{Tr} (\partial_m T)>0$ for any real
fields $\{A_t\}_{t=0}^{N-1}$, the sign problem in this
representation is avoided.  These averages $ \langle \cdots \rangle_0$ can be calculated by the
usual Monte Carlo method or Langevin dynamics since these averages
is taken over the Gaussian weight $\exp
(-\frac{1}{8U}\sum_{t=0}^{N-1} A_t^2 )$. We use (real) Langevin
dynamics
\begin{eqnarray}\label{2016_5_4_11_1}
 A_{t,\Theta+\Delta \Theta} = A_{t, \Theta}   -\Delta \Theta  \frac{1}{4U}A_t + \sqrt{2\Delta \Theta} \eta_{t,\Theta}, \quad 0\leq
t\leq N-1, \quad \Theta = 0, 1,\cdots
\end{eqnarray}
 to calculate these averages. Here $\Theta$ denotes the discrete Langevin time and $\Delta \Theta$
is the time step. The real white noise $\eta_{t,\Theta}\sim {\cal N}
(0,1)$ satisfies $ \langle \eta_{t,\Theta} \eta_{\tau,\Theta^\prime}
\rangle = \delta_{t,\tau}\delta_{\Theta,\Theta^\prime}$. The
averages $\langle \textrm{Tr} (T) \rangle_0$ and $\langle
\textrm{Tr} (\partial_m T) \rangle_0$, which does not depend on the
chemical potential $\mu$, are obtained by samples
$\{A_t\}_{t=0}^{N-1}$ after the equilibrium of the real Langevin
dynamics. Moreover, the chiral condensate in (\ref{2016_5_4_10}) and
fermion density in (\ref{2016_5_4_11}) can be obtained from $\langle
\textrm{Tr} (T) \rangle_0$ and $\langle \textrm{Tr} (\partial_m T)
\rangle_0$ for all chemical potential $\mu$.

 The computational complexity
for the calculation of $\textrm{Tr} ( T)$ and $\textrm{Tr}
(\partial_m T)$ are $O(N)$, which will be explained in the next
section. The exact formula for $Z_2(m)$ is known
\cite{Pawlowski_2013_094503},
\begin{eqnarray}\label{2016_5_4_24}
Z_2(m) = 2(2\pi U)^{N/2}(B_++B_-)
\end{eqnarray}
\begin{eqnarray}\label{2016_5_4_24_1}
  B_\pm =
\frac{m^N}{2}\Big(1\pm \sqrt{4UB_c}\Big)^{N}, \quad B_c =
\frac{1}{4U} + (\frac{1}{4U}+1)\frac{1}{m^2}
\end{eqnarray}

Thus these averages can be calculated exactly, which is called the
exact result in the following sections.

\section{Numerical results}\label{results}
\subsection{Implementation}
The most advantage of the fermion bag approach is that the weight in
(\ref{2016_4_27_2}) is nonnegative and thus the sign problem
is avoided in the interesting range of parameters, $0\leq \mu\leq 2$,
$0<m<100$, $0\leq U\leq 10$.  Table \ref{tab1} shows
the values of $C(t_1,\cdots,t_j)$ for randomly chosen $j$ sites
$t_1,\cdots,t_j$.

The Monte Carlo implementation for the fermion bag approach,
including three steps of configuration updating and sampling of
chiral condensate and fermion density, depends on the ratio of the
value of $C$.  For small coupling constant $U$, there are few bonds
connecting neighboring sites, we use the propagation matrix on the
occupied sites to calculate this ratio; For large coupling constant
$U$, many bonds connecting to neighboring sites appear, we use the
fermion matrix on the free sites to calculate this ratio. Thus high
computational efficiency is achieved for the fermion bag approach
with weak or strong coupling between fermions. In our code, we use
$\textrm{D\_org}$ and $\textrm{D\_inv}$ to denote the fermion matrix
$D$ and its inverse matrix, respectively, which does not change in
the whole run. $\textrm{free\_sites}=\backslash
\textrm{occu\_sites}$ and $\textrm{occu\_sites}$ denote the free
sites and occupied sites for the current configuration $k$. Denote
by $\textrm{D\_Inv[free\_sites]}$ the inverse of
$\textrm{D\_org[free\_sites]}$ and by $\textrm{D\_Org[occu\_sites]}$
the inverse of $\textrm{D\_inv[occu\_sites]}$, respectively. Here
$\textrm{D\_org[free\_sites]}$ ($\textrm{D\_inv[occu\_sites]}$)
denotes the submatrix of $\textrm{D\_org}$ ($\textrm{D\_inv}$) with
the rows and columns corresponding to $\textrm{free\_sites}$
($\textrm{occu\_sites}$). Thus depending on the number of occupied
sites,  $\textrm{D\_Inv}$ or $\textrm{D\_Org}$ changes during Monte
Carlo updating.

The initial configuration for $k$ is the one where there are no
bonds. The sampling is taken after $1\times 10^6$ Monte Carlo steps,
where in each step at most two bonds change. To reduce the
autocorrelation effects, two subsequent samples are separated by
$10N$ Monte Carlo steps.  The sampling is finished after $1\times
10^7$ Monte Carlo steps, thus about $9\times 10^6/(10N)$ samples are
taken for each Monte Carlo simulation.

The implementation of the complex Langevin dynamics
(\ref{2016_5_2_1}) is rather simple. The initial condition for $A$
is zero everywhere. We choose the time step $\Delta \Theta =
10^{-4}$. The sampling is taken after $t_{\textrm{equ}}=20$ (i.e.,
$20/\Delta \Theta=2\times 10^5$ complex Langevin steps). To compare
with the fermion bag approach, two subsequent samples are separated
by 10 complex Langevin steps. The end time in the complex Langevin
dynamics is $t_{\textrm{end}}=40$, thus about $20\times 10^6/(10N)$
samples are taken for each complex Langevin simulation.

To calculate the sampling (\ref{2016_5_4_10}) with the computational
complexity $O(N)$, we introduce
$$ C_i = B_0\cdots B_i, \quad i=0,\cdots,N-1, \quad
D_i = B_{i+1}\cdots B_{N-1}, \quad i=0,\cdots,N-2 $$ where $B_i =
\left(
\begin{array}{cc}
m &  \frac{1}{4}(1+A_{i}^2) \\
1 & 0
\end{array} \right)$, $i=0,\cdots,N-1$. Using the iteration
$$C_{i+1}=C_iB_{i+1}, \quad i=0,\cdots,N-2, \quad D_{i}=B_{i+1}D_{i+1}, \quad i=N-3,\cdots,0$$
we know that the computational complexity for calculating
$\{C_i\}_{i=0}^{N-1}$ and $\{D_i\}_{i=0}^{N-2}$ is $O(N)$. The trace
of $T$ is $\textrm{Tr}(T)=\textrm{Tr}(C_{N-1})$ and
$$  \textrm{Tr} ( \partial_m
T ) = \textrm{Tr} (FD_0) + \sum_{i=1}^{N-2} \textrm{Tr} (FD_i
C_{i-1}) + \textrm{Tr} (FC_{N-2}), \quad F = \left(
\begin{array}{cc}
1 &  0 \\
0 & 0
\end{array} \right)$$

We use the $\Gamma$ method to estimate the error for the samples in
each Monte Carlo simulation or complex Langevin dynamics
\cite{Wolff.143}. A typical result is shown in Figure \ref{fig1}
where the statistical error for the complex Langvin dynamics are
larger than those for the fermion bag approach if the chemical
potential is close to 1.

\begin{table}\label{tab1}
\centering
\begin{tabular}{c|c}
\hline
$t_1,\cdots,t_j$ & $C(t_1,\cdots,t_j)$  \\
\hline
empty    &            11.65480 \\
 3        &       0.006890  \\
 3 6 & 0.018226 \\
3 4 5 & 0.019760 \\
1 2 6 7 &         0.005100  \\
0 2 3 6 7 & 0.026000 \\
0 1 2 4 6 7 & 0.010000 \\
 0 1 2 3 4 5 7 &   0.100000 \\
  0 1 2 3 4 5 6 7 & 1.000000 \\
\end{tabular}
\caption{ The value of $C$ for randomly chosen occupied sites
$(t_1,\cdots,t_j)$, $m=0.1$, $\mu=1.0$, $N=8$.} \label{tab1}
\end{table}

\begin{figure}
\centering
\includegraphics[width=8cm,height=6cm]{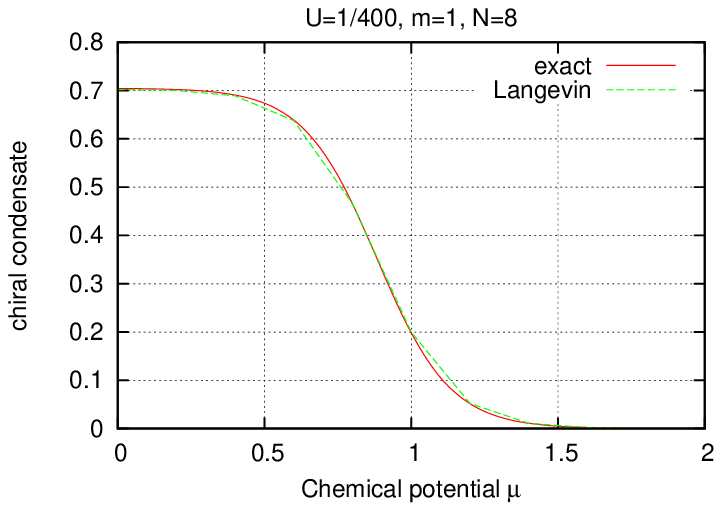}
\includegraphics[width=8cm,height=6cm]{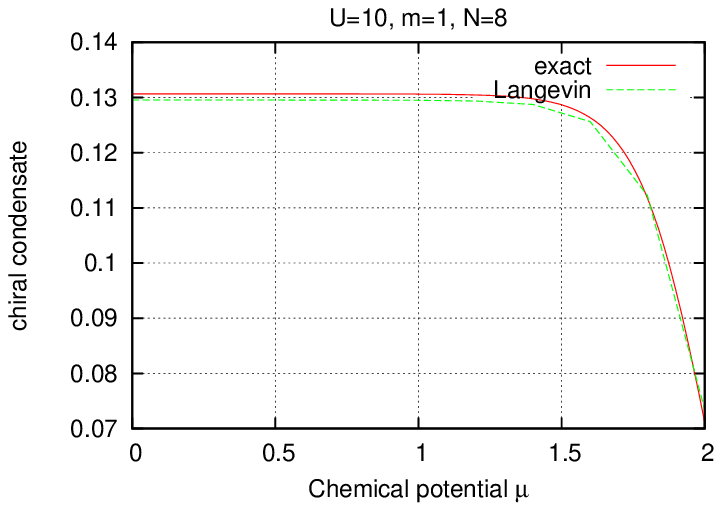}
\caption{Comparison between Langevin dynamics and exact solution.
Upper: $\Delta\Theta=0.01$, $t_{\textrm{eq}}=800$,
$t_{\textrm{end}}=1600$, $\langle \textrm{Tr} (T)\rangle_0=4.6145\pm
0.000608$ and $\frac{1}{N}\langle \textrm{Tr} (\partial_m
T)\rangle_0=3.2466\pm 0.000369$. Bottom: $\Delta\Theta=0.025$,
$t_{\textrm{eq}}=30000$, $t_{\textrm{end}}=50000$,  $\langle
\textrm{Tr} (T)\rangle_0= 45457.8 \pm  4456.4  $ and
$\frac{1}{N}\langle \textrm{Tr} (\partial_m T)\rangle_0= 5888.6 \pm
405.1$. }\label{fig0}
\end{figure}

\subsection{Simulation results}
 The exact result
is known for the massive Thirring model with one flavor in 0+1
dimension \cite{Pawlowski_2013_094503}.  We first compare the chiral
condensate obtained by the Langevin dynamics and exact result. Since
the sign problem is avoided and the observable is positive, the
Langvin dynamics (See Eq.
(\ref{2016_5_4_10})(\ref{2016_5_4_11})(\ref{2016_5_4_11_1})) should
reproduce the exact result. Figure \ref{fig0} shows the agreement
between the chiral condensate obtained by the Langvin dynamics and
exact results. In fact, the agreement between them are also achieved
for the parameters: $m=0.1,1,3,100$, $U=1/400,1/12,1/4,10$,
$\mu=0,0.2,0.4,\cdots,2.0$.

\begin{figure}
\centering
\includegraphics[width=8cm,height=6cm]{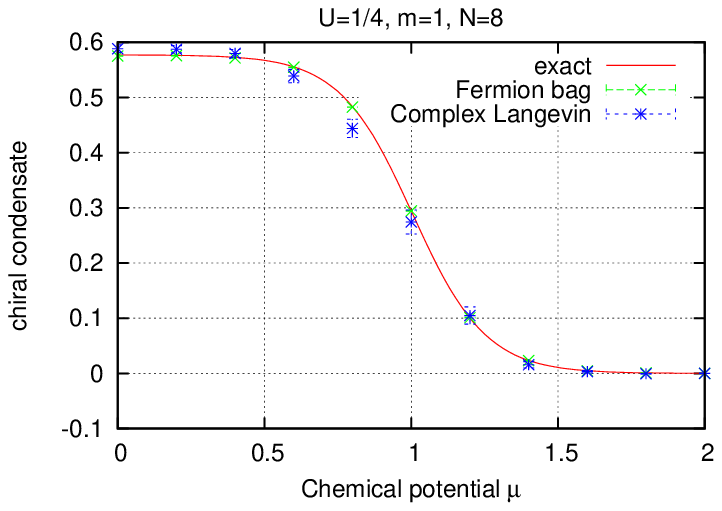}
\includegraphics[width=8cm,height=6cm]{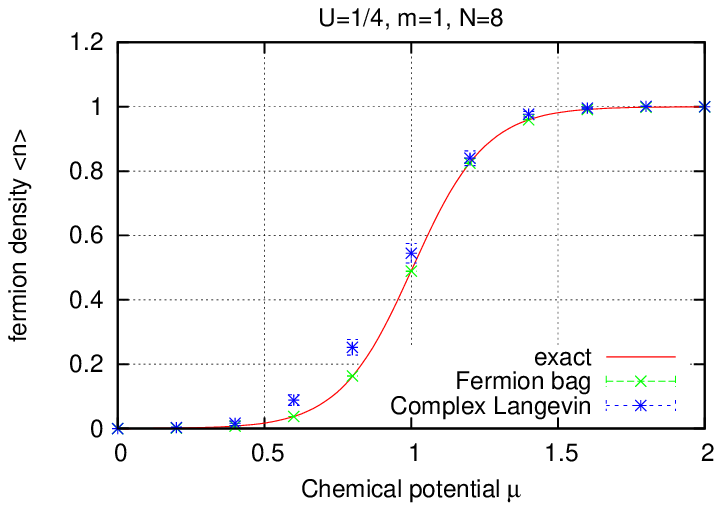}
\caption{Comparison between Fermion bag approach, complex Langevin
dynamics and exact solution.  }\label{fig1}
\end{figure}

\begin{figure}
\centering
\includegraphics[width=8cm,height=6cm]{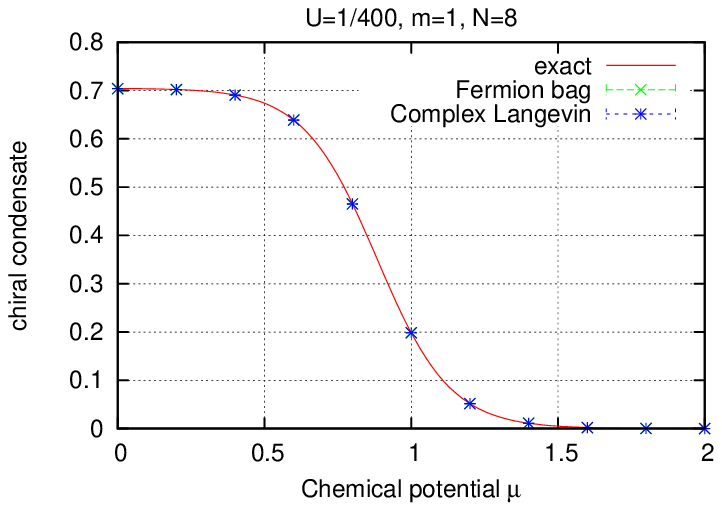}
\includegraphics[width=8cm,height=6cm]{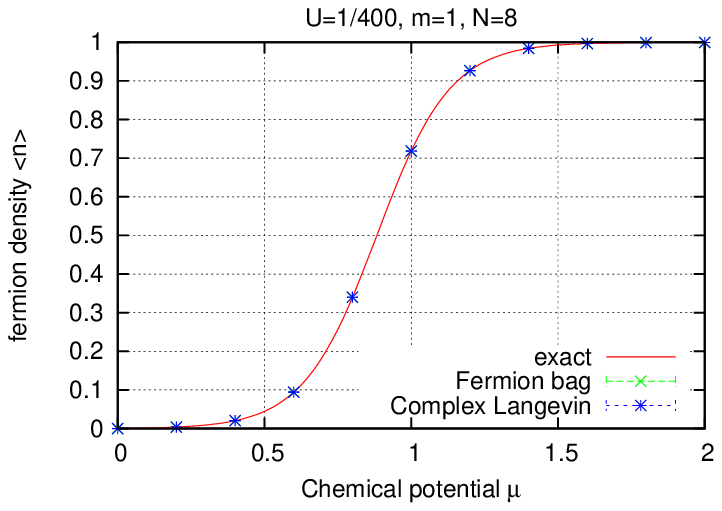}
\caption{Comparison between Fermion bag approach, complex Langevin
dynamics and exact solution.  }\label{fig2}
\end{figure}

\begin{figure}
\centering
\includegraphics[width=8cm,height=6cm]{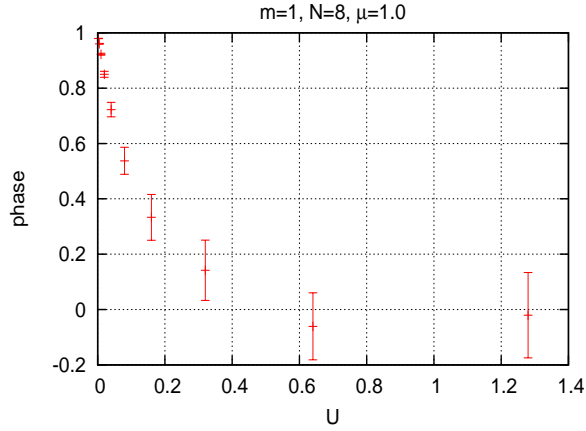}
\caption{The dependence of phase $\langle e^{i\varphi}\rangle = \frac{Z}{Z_{\text{pq}}}$ on the fermion coupling strength
$U=0.0025,0.005,0.01,0.02,0.04,0.08,0.16,0.32,0.64,1.28$, where $Z_{\text{pq}}$ is the phase quenched approximation
  of $Z$ in (\ref{2016_4_27_10}), $\Delta \Theta=10^{-4}$, $t_{\text{equ}}=40$, $t_{\text{end}}=80$.  }\label{fig10}
\end{figure}

 Figure \ref{fig1}
shows the comparison between fermion bag approach, complex Langevin
dynamics with the exact results. The chiral condensate and fermion
density for fermion bag approach agree very well with the exact
result in the range $0\leq \mu\leq 2$. Compared with the fermion bag
approach, the result obtain by complex Langevin dynamics agree with
exact result for small or large $\mu$. Moreover the statistical
error becomes large for the complex Langevin dynamics in the
intermediate value $\mu$. Figure \ref{fig2} shows that the fermion
bag approach and complex Langevin dynamics agree with the exact
results very well when $U$ is decreased to be $U=1/400$. Since the
interaction between fermions is decreased, the statistical error is
also small compared with those in Figure \ref{fig1}. When $U$ increased, the simulation results obtained by complex Langevin dynamics
is not reliable. According to Ref. \cite{Aarts_2011_3270}\cite{Pawlowski_2013_094503}, the quantity
$$ \sum_{t=1}^N \Big(\frac{d}{dA_t} - \frac{dS_{\text{eff}}}{dA_t}\Big)\frac{d}{dA_t} O(A)  $$
should vanish for any holomorphic function $O(A)$ if the complex Langevin dynamics works. We choose the observable (the chiral condensate) $O(A)=\frac{1}{N}\text{Tr}(K^{-1})$.  For $\mu=1$, $m=1$ and $N=8$, it is $0.0137\pm 0.00708$ which is
  small if $U=0.0025$ and becomes $-5.88\pm 7.33$  if $U=0.16$. Moreover, the statistical error is also large which means that
  it is difficult to measure it reliably when $U$ is increased. Figure \ref{fig10} shows the dependence of the phase $\langle e^{i\varphi}\rangle
  _{\text{pq}}
  =Z/Z_{\text{pq}}$ on the coupling strength $U$ for fixed chemical potential $\mu=1$. Here $Z_{\text{pq}}$ is the phase quenched approximation of $Z$ in (\ref{2016_4_27_10}) where $\det K$ is replaced by its module
  $|\det K|$. When $U$ is increased the phase $\langle e^{i\varphi}\rangle
  _{\text{pq}}$ decays to zero, which shows it suffers from a severe sign problem. Moreover, the statistic error
  is also increased if $U$ becomes larger.

\begin{figure}
\centering
\includegraphics[width=8cm,height=6cm]{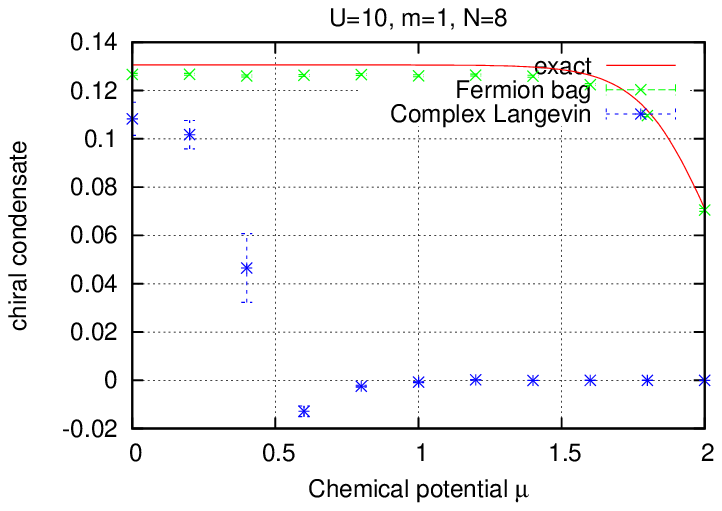}
\includegraphics[width=8cm,height=6cm]{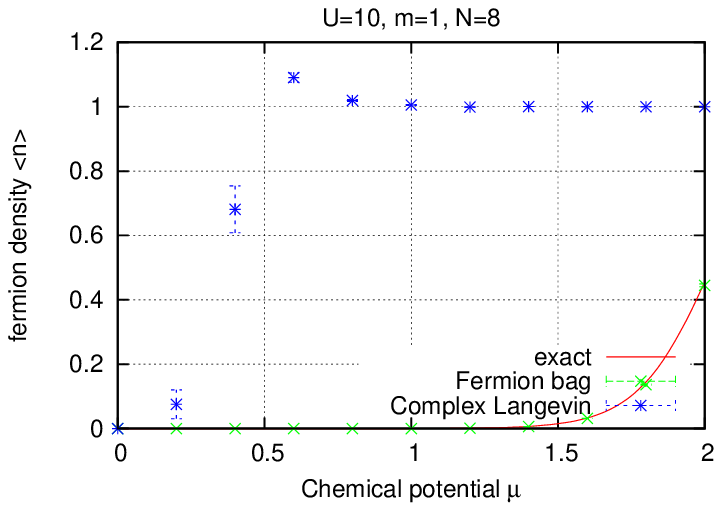}
\caption{Comparison between Fermion bag approach, complex Langevin
dynamics and exact solution.  }\label{fig3}
\end{figure}

\begin{figure}
\centering
\includegraphics[width=8cm,height=6cm]{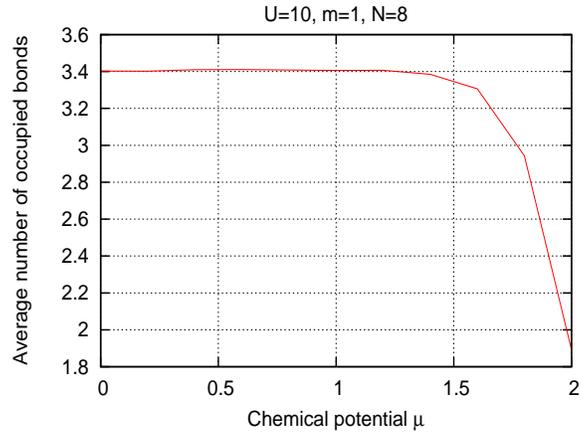}
\caption{The average number of occupied bonds vs the chemical
potential $\mu$ by fermion bag approach. }\label{fig5}
\end{figure}

\begin{figure}
\centering
\includegraphics[width=6cm,height=4cm]{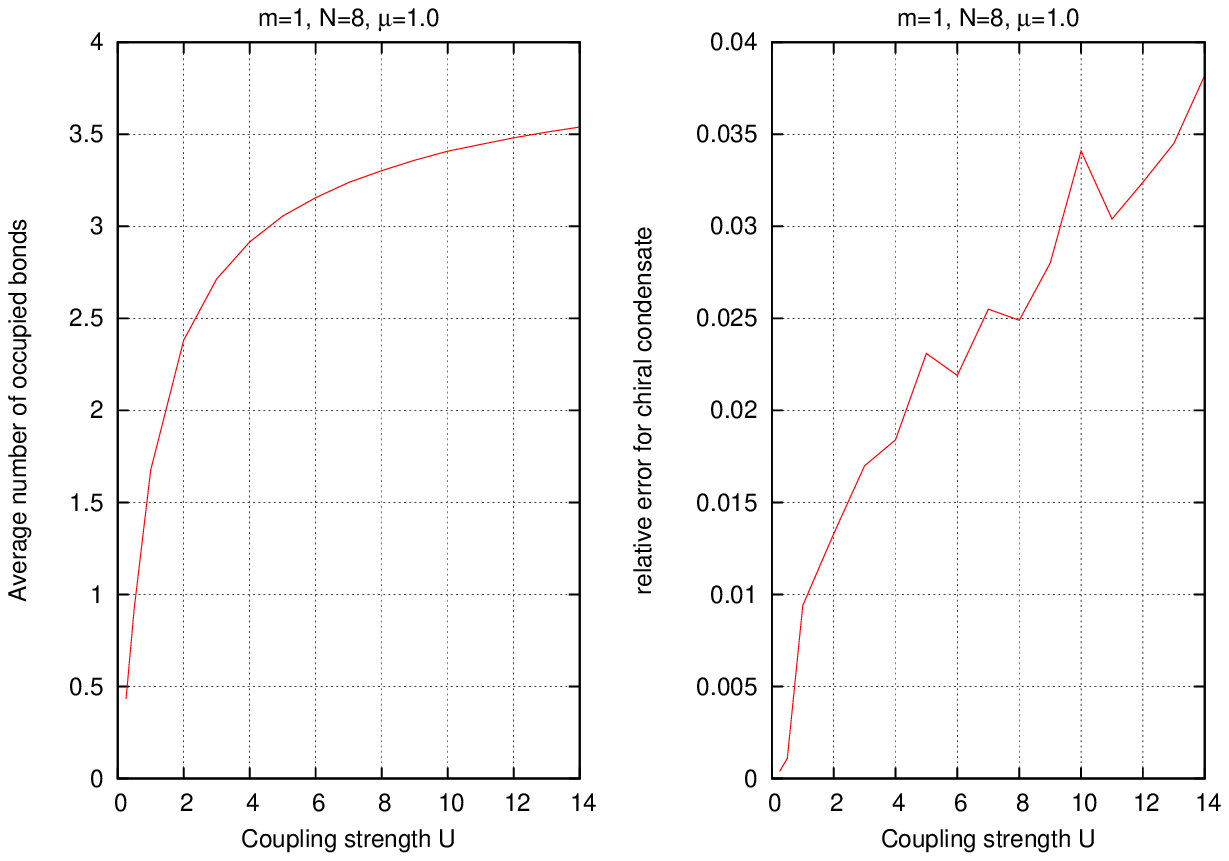}
\caption{The average number of occupied bonds and relative error for
chiral condensate vs the coupling strength $U$ by fermion bag
approach. }\label{fig6}
\end{figure}

When $U=10$, the results obtained by the fermion bag approach agree
with exact result rather well, but the complex Langevin dynamics is
totally wrong, which is shown in Figure \ref{fig3}. The failure of
the complex Langevin dynamics for strong interaction was also found
in Ref. \cite{Pawlowski_2013_094503}. For $\mu=1.0$ in Figure
\ref{fig3}, the average number of occupied bonds in the fermion bag
approach is 3.41, which means almost all
 sites are occupied by bonds $(N=8)$. This leads to a slight
 difference of the chiral condensate obtained by the fermion bag
 approach and exact results. Figure \ref{fig5} show that the average number of occupied bonds drops rapidly near $\mu=2$ where the simulated results
 obtained by fermion bag approach agree with the exact result. Here we don't show the error bar in Figure \ref{fig5} since it is very small which can be shown for the error bar of the chiral condensate by fermion bag approach in Figure
 \ref{fig3}. For fixed $m=1$ and $\mu=1$, \ref{fig6} shows the dependence of the average number of occupied bonds
 (Left) and relative error for chiral condensate (Right) on the
 coupling strength $U$. The average number of occupied bonds
 increase with the coupling strength $U$. Since the exact solution
 is known, the relative error for the chiral condensate (obtained by
 the fermion bag approach) can be calculated as shown in the right
 figure of Figure \ref{fig6} where the relative error is increased
with the coupling strength when $U<5$ or $U>11$. We also notice
there is an oscillation of the relative error for $6<U<11$. Thus if
the coupling strength $U$ is not too large the fermion bag approach
can always reproduce the exact results. When $U$ is very large,
almost all lattice sites are occupied, which will lead a slight
difference between results obtained by fermion bag approach and the
exact results. It should be mentioned that the fluctuation of the
chiral condensate and fermion density in the fermion bag approach is
rather small even when $U$ is rather large, e.g., $10\leq U\leq 14$.

   For the parameter: $U=10$, $m=1$ and $N=8$, the chiral
condensate obtained by Langevin dynamics in Figure \ref{fig0}
(bottom) and those by fermion bag approach in Figure \ref{fig3}
(upper) can reproduce the exact result. Compared with the fermion
bag approach, the equilibrium time in the Langevin dynamics is very
long. However, the advantage of the Langevin dynamics over the other
two approaches is that the chiral condensate and fermion density for
any chemical potential can be obtained when the averages $\langle
\textrm{Tr} (\partial_m T)\rangle_0$ and $\langle \textrm{Tr}
(T)\rangle_0$  is
calculated from the Langevin dynamics, since they does not depend on the chemical potential.

The comparison are also made between the fermion bag approach,
complex Langevin dynamics with exact results in the range of
parameters: $m=0.1,1,3,100$, $U=1/400,1/12,1/4,10$,
$\mu=0,0.2,0.4,\cdots,2.0$. The chiral condensate and fermion
density obtained by the fermion bag approach agrees very well with
the exact results for all these parameters. The reason is that the
weight in the fermion bag approach is nonnegative and thus the sign
problem is avoided. The results obtained by complex Langevin
dynamics agree with the exact results only if $U$ is not too large,
$U<O(1)$. When $U$ is increased, compared with the drift term, the
white noise fluctuation in complex Langevin equation becomes large
and thus the complex Langevin does not work, especially, when the
fermion mass $m$, e.g., $m=0.1$, is rather small.

\section{Conclusions}\label{conclusion}
The fermion bag approach, Langevin dynamics and complex Langevin
dynamics are used to solve the massive Thirring model at finite
density in 0+1 dimension.  The chiral condensate and fermion density
by these methods are compared with the exact results. The complex
Langevin dynamics only works for not too large interaction $U<O(1)$.
Moreover, this method will also meet the computational difficulties
when the fermion mass is too small, or the chemical potential is in
the intermediate range $\mu=1$.  Since the sign problem is avoided
in the Langevin dynamics and the fermion bag approach, these two
methods can reproduces the exact results for a large range of
parameters. Another advantage of the fermion bag approach over the
Langevin dynamics is that a high computational efficiency are made
for the strong interaction between fermions for the fermion bag
approach. In the future paper we will check these advantages of the
fermion bag approach over the other numerical methods for the
massive Thirring model at finite density in 2+1 dimensions.

\vspace{1cm}

 Acknowledgments.
  I would like to thank Prof. Shailesh
Chandrasekharan for discussion. Daming Li was supported by the
National Science Foundation of China (No. 11271258, 11571234).

{}

\appendix

\section{\label{Appendix_1}There is no sign problem for the presentation of $Z$ by fermion bag approach in (\ref{2016_4_27_2})}
The $N\times N$ (even $N$) fermion matrix is
\begin{eqnarray*}
D = D(\mu,m) = \left( \begin{array}{cccccc}
m  &  \frac{e^\mu}{2} &  &  &  & \frac{e^{-\mu}}{2}  \\
-\frac{e^{-\mu}}{2} & m  &  \frac{e^{\mu}}{2}   &  &  & \\
& -\frac{e^{-\mu}}{2} & m  &  \frac{e^{\mu}}{2}    &  &  \\
&  &   &   \ddots   &  &  \\
    &  &   &  & m &  \frac{e^{\mu}}{2} \\
-\frac{e^{\mu}}{2}  &   & &   & -\frac{e^{-\mu}}{2}  & m   \\
\end{array} \right)_{N\times N}
\end{eqnarray*}
According to a formula of the determinant \cite{Molinari.2221}, the
determinant of $D$ is
\begin{eqnarray*}
\det D & =&      \frac{e^{N\mu}}{2^N}  + \frac{e^{-N\mu}}{2^N}  +
\textrm{Tr} (T)
\end{eqnarray*}
where the $2\times 2$ transfer matrix $T$ is $ T  =  \left(
\begin{array}{cc}
m &  \frac{1}{4}  \\
1 & 0
\end{array} \right)^N
$. Obviously, $\det D>0$ for any $\mu>0$ and $m>0$. Choose $n$
different indices, $1\leq i_1<\cdots<i_n\leq N$ and  delete $n$ rows
and columns corresponding to these $n$ indices from $D$ to obtain
$\tilde D$. We want to prove that $(N-n)\times (N-n)$ matrix $\tilde
D$ satisfies $\det \tilde D>0$. This holds because the structure of
$\tilde D$ is the same with $D$ and thus the determinant of $\tilde
D$ can be calculated \cite{Molinari.2221}, which must be positive.
For example, $N=10$, $n=2$, $i_1=4$, $i_2=7$,
\begin{eqnarray*}
\tilde D =   \left( \begin{array}{ccccccccccc}
*  &  * &  & | & & & | & &  & * \\
* &  *  & * & | &  & & | & &  \\
& * &  *  &  | &  &  & | &  & & \\
-& - &  -  & - & - & - & - & - & - & - \\
 &  &    & |  & *  & * & | &  &  &  \\
 &  &    & |  & *  & * & | &  &  &  \\
-& - &  -  & - & - & - & - & - & - & - \\
&  &    & | &  &  & | & * & * &  \\
&  &    & | &  &  & | & * & * & * \\
* &  &    & | &  &  & | &  & * & * \\
\end{array} \right)
\end{eqnarray*}
Since $C(t_1,\cdots,t_{2j})$ can be presented by the determinant of
the submatrix of $D$, which is nonnegative, the sign problem is
avoided for the presentation of $Z$ by fermion bag approach in (\ref{2016_4_27_2}).

\end{document}